\documentclass[11pt]{article}

\usepackage[margin=1in]{geometry}
\usepackage{amsmath,amssymb,amsthm}
\usepackage{graphicx}
\usepackage{booktabs}
\usepackage{caption}
\usepackage{subcaption}
\usepackage{cite}
\usepackage{xcolor}
\usepackage{indentfirst}
\usepackage{multirow}
\usepackage{pgfplots}
\usepackage{pgfplotstable}
\usepackage{tikz}
\pgfplotsset{compat=1.14}
\usepackage{hyperref}

\theoremstyle{definition}
\newtheorem{definition}{Definition}
\newtheorem{theorem}{Theorem}
\newtheorem{proposition}{Proposition}

\newtheorem{example}{Example}
\newtheorem{remark}{Remark}

\newcommand{\Stab}{\mathcal{S}}
\newcommand{\Pauli}{\mathcal{P}}
\newcommand{\ket}[1]{|#1\rangle}

\title{Distributed Hyperbolic Floquet Codes under Depolarizing and Erasure Noise}
\date{}
\author{Aygul Azatovna Galimova\textsuperscript{1}\\[4pt]
\textsuperscript{1}\textit{Department of Mathematics}\\
\textit{Duke University, Durham, NC 27708, USA}}

\begin{document}

\maketitle

\begin{abstract}
Distributing qubits across quantum processing units (QPUs) connected by shared entanglement enables scaling beyond monolithic architectures. Hyperbolic Floquet codes use only weight-2 measurements and are good candidates for distributed quantum error correcting codes.

We construct hyperbolic and semi-hyperbolic Floquet codes from $\{8,3\}$, $\{10,3\}$, and $\{12,3\}$ tessellations via the Wythoff kaleidoscopic construction with the Low-Index Normal Subgroups (LINS) algorithm and distribute them across QPUs via spectral bisection. The $\{10,3\}$ and $\{12,3\}$ families are new to hyperbolic Floquet codes.

We simulate these distributed codes under four noise models: depolarizing, SDEM3, correlated EM3, and erasure. With depolarizing noise ($p_{\text{local}} = 0.03\%$), fine-grained codes achieve non-local pseudo-thresholds up to 3.0\% for $\{8,3\}$, 3.0\% for $\{10,3\}$, and 1.75\% for $\{12,3\}$. Correlated EM3 yields pseudo-thresholds up to 0.75\% for $\{8,3\}$, 0.75\% for $\{10,3\}$, and 0.50\% for $\{12,3\}$; crossing-based thresholds from same-$k$ families are ${\sim}1.75$--$2.9\%$ across all tessellations. Using the SDEM3 model, fine-grained codes achieve distributed pseudo-thresholds of 1.75\% for $\{8,3\}$, 1.25\% for $\{10,3\}$, and 1.00\% for $\{12,3\}$. Under erasure noise motivated by spin-optical architectures, thresholds at 1\% local loss are 35--40\% for $\{8,3\}$, 30--35\% for $\{10,3\}$, and 25--30\% for $\{12,3\}$.
\end{abstract}

\tableofcontents
\newpage

\section{Introduction}
\label{sec:introduction}

\indent Quantum error correction is necessary for fault-tolerant quantum computation~\cite{nielsen2010quantum, shor1995scheme}. Topological codes such as the surface code~\cite{kitaev2003fault, dennis2002topological} are widely studied for near-term quantum error correction due to their planar geometry and nearest-neighbor connectivity~\cite{fowler2012surface}.

\indent A limitation of planar surface codes is their encoding rate: the ratio $k/n$ of logical to physical qubits approaches zero as code size increases~\cite{bravyi2010tradeoffs}. Hyperbolic surface codes achieve constant rate $k/n = \Theta(1)$ by embedding qubits on negatively curved surfaces~\cite{breuckmann2016constructions}. However, their face stabilizers have weight $p$ (e.g., weight 8 for $\{8,3\}$)~\cite{breuckmann2016constructions}.

\indent Floquet codes~\cite{hastings2021dynamically} replace high-weight stabilizers with periodic sequences of weight-2 measurements that dynamically generate the code space. The honeycomb Floquet code~\cite{gidney2021honeycomb} achieves circuit-level thresholds comparable to the surface code. Fahimniya et al.~\cite{fahimniya2023faulttolerant} and Higgott and Breuckmann~\cite{higgott2024constructions} introduce $\{8,3\}$ hyperbolic Floquet codes. Higgott and Breuckmann present semi-hyperbolic variants with improved distance scaling. Ozawa et al.~\cite{ozawa2025hyperbolic} propose a simplified schedule using only $XX$ and $ZZ$ measurements.

\indent Scaling quantum computers beyond the thousands of physical qubits available on a single chip will likely require distributing qubits across networked quantum processing units (QPUs), with inter-QPU operations mediated by shared entanglement. Ramette et al.~\cite{ramette2024faulttolerant} use Greenberger--Horne--Zeilinger (GHZ) states to perform weight-4 stabilizer measurements across QPU boundaries in a distributed surface code. Sutcliffe et al.~\cite{sutcliffe2025distributed} evaluate hyperbolic Floquet codes in a distributed architecture; their largest semi-hyperbolic $\{8,3\}$ code achieves a non-local depolarizing pseudo-threshold of ${\sim}3.1\%$.

\indent We construct hyperbolic and semi-hyperbolic Floquet codes from $\{8,3\}$, $\{10,3\}$, and $\{12,3\}$ tessellations using the LINS algorithm with the Wythoff kaleidoscopic construction and evaluate them in distributed architectures. Prior work on hyperbolic Floquet codes has focused on $\{8,3\}$ tessellations~\cite{fahimniya2023faulttolerant, higgott2024constructions, ozawa2025hyperbolic, sutcliffe2025distributed}; the $\{10,3\}$ and $\{12,3\}$ families are new. We test four noise models:
\begin{enumerate}
\item Distributed depolarizing noise (Section~\ref{sec:distributed}).
\item SDEM3 noise (Section~\ref{sec:sdem3_distributed}).
\item Correlated EM3 noise (Section~\ref{sec:correlated_em3_distributed}).
\item Erasure (Section~\ref{sec:distributed_erasure}).
\end{enumerate}

\section{Stabilizer Codes and Floquet Codes}

\subsection{The Stabilizer Formalism}

\indent The stabilizer formalism~\cite{gottesman1997stabilizer} provides a group-theoretic framework for quantum error correction. For $n$ qubits, the Pauli group is $\Pauli_n = \{\pm 1, \pm i\} \times \{I, X, Y, Z\}^{\otimes n}$.

\begin{definition}[Stabilizer Code]
An $[[n, k, d]]$ stabilizer code is a $2^k$-dimensional subspace of $(\mathbb{C}^2)^{\otimes n}$ determined by an abelian subgroup $\Stab \subseteq \Pauli_n$:
\begin{equation}
\mathcal{C} = \{\ket{\psi} \in (\mathbb{C}^2)^{\otimes n} : g\ket{\psi} = \ket{\psi} \text{ for all } g \in \Stab\},
\end{equation}
where $\Stab$ does not contain $-I$, $|\Stab| = 2^{n-k}$, and $\Stab = \langle g_1, \ldots, g_{n-k} \rangle$.
\end{definition}

\begin{definition}[Logical Operators]
Logical operators are elements of the normalizer modulo the stabilizer:
\begin{equation}
\text{Logicals} = N(\Stab) / \Stab,
\end{equation}
where $N(\Stab) = \{P \in \Pauli_n : PgP^{-1} \in \Stab \text{ for all } g \in \Stab\}$.
\end{definition}

\begin{definition}[Code Distance]
The distance of a stabilizer code is:
\begin{equation}
d = \min\{\text{wt}(L) : L \in N(\Stab) \setminus \Stab\},
\end{equation}
where $\text{wt}(L)$ is the number of non-identity tensor factors of the Pauli operator $L$.
\end{definition}

\indent For Floquet codes, the code space is dynamically generated, and the relevant distance metric is the \emph{embedded distance} $d_{\text{emb}}$~\cite{higgott2024constructions}, defined in Section~\ref{sec:code_construction}.

\begin{definition}[CSS Code]
A Calderbank-Shor-Steane (CSS) code is a stabilizer code whose generators partition into $X$-type (products of $X$ only) and $Z$-type (products of $Z$ only).
\end{definition}

\subsection{Surface Code Stabilizers}

\indent For CSS surface codes on tessellations, qubits are placed on edges with stabilizers associated with faces and vertices. We write $\Stab$ for the stabilizer group and use $S_f$, $S_v$ for individual stabilizer operators (elements of $\Stab$).

\begin{definition}[Plaquette and Vertex Stabilizers]
\label{def:plaquette_vertex}
For each face $f$, define $S_f^Z = \prod_{e \in \partial f} Z_e$. For each vertex $v$, define $S_v^X = \prod_{e \in \text{star}(v)} X_e$.
\end{definition}

\indent For hyperbolic Floquet codes, the qubit placement differs: physical qubits are placed on vertices, and two-body measurements act on edges (pairs of adjacent vertices). Each edge connects exactly two vertices, so $XX$, $YY$, or $ZZ$ measurements act between adjacent qubits.

\begin{proposition}[Number of Logical Qubits]
For a surface code on a genus-$g$ surface with $V$ vertices, $E$ edges, and $F$ faces:
\begin{equation}
k = E - (F-1) - (V-1) = E - F - V + 2 = 2g,
\end{equation}
using the Euler characteristic $\chi = V - E + F = 2 - 2g$. The $F$ face stabilizers satisfy one linear dependency ($\prod_f S_f^Z = I$) and therefore have rank $F - 1$. The $V$ vertex stabilizers satisfy one dependency ($\prod_v S_v^X = I$) and have rank $V - 1$~\cite{breuckmann2016constructions}.
\end{proposition}

\subsection{The Floquet Measurement Cycle}

\indent A Floquet code~\cite{hastings2021dynamically} is defined by a periodic sequence of non-commuting two-body measurements that dynamically generate a code space~\cite{fahimniya2023faulttolerant}. In each round, a set of commuting weight-2 Pauli operators is measured; operators in consecutive rounds generally anti-commute. Our $\{p,3\}$ tessellations are 3-edge-colorable (Section~\ref{sec:face_coloring}), so we adopt the $XX/YY/ZZ$ schedule~\cite{fahimniya2023faulttolerant, higgott2024constructions}, which cycles through three rounds:
\begin{itemize}
\item Round 0 (Red): Measure $XX$ on all red edges.
\item Round 1 (Green): Measure $YY$ on all green edges.
\item Round 2 (Blue): Measure $ZZ$ on all blue edges.
\end{itemize}

\indent The \emph{instantaneous stabilizer group} (ISG) is the stabilizer group of the code at a given point in the schedule~\cite{hastings2021dynamically}. Each round updates the ISG: the new check measurements replace anti-commuting generators from the previous round. After reaching steady state, the state is in the $+1$-eigenspace of all plaquette stabilizers~\cite{higgott2024constructions}.

\begin{definition}[Floquet Face Stabilizer]
\label{def:floquet_face}
The face stabilizer $S_f$ for a $p$-gon face $f$ is the product of the $p$ edge checks around the boundary:
\begin{equation}
S_f = \prod_{e \in \partial f} M_e,
\end{equation}
where $M_e \in \{XX, YY, ZZ\}$ depends on the edge color.
\end{definition}

\subsection{Weight-2 Decomposition}

\indent High-weight face stabilizers can be inferred from products of weight-2 edge measurements. Consider a hexagonal face with alternating edge colors R, G, B, R, G, B. Writing $M_{e_i}$ for the measurement on edge $e_i$ as in Definition~\ref{def:floquet_face}, the face stabilizer is:
\begin{equation}
S_f = M_{e_1} \cdot M_{e_2} \cdot M_{e_3} \cdot M_{e_4} \cdot M_{e_5} \cdot M_{e_6} = (XX)_1 \cdot (YY)_2 \cdot (ZZ)_3 \cdot (XX)_4 \cdot (YY)_5 \cdot (ZZ)_6,
\end{equation}
where indices label edges around the face. Each vertex is incident to exactly two edges with different colors. Two different Paulis therefore act on each vertex. The product simplifies using $XY = iZ$, $YZ = iX$, $ZX = iY$.

\indent Detectors are constructed by comparing plaquette measurement outcomes across consecutive cycles~\cite{higgott2024constructions}.

\section{Hyperbolic Geometry and Tessellations}

\subsection{The Poincar\'e Disk Model}

\indent The hyperbolic plane $\mathbb{H}^2$ is a two-dimensional Riemannian manifold with constant negative Gaussian curvature $K = -1$. In the Poincar\'e disk model $\mathbb{D} = \{z \in \mathbb{C} : |z| < 1\}$, the metric is:
\begin{equation}
ds^2 = \frac{4|dz|^2}{(1-|z|^2)^2}.
\end{equation}

\indent The distance between two points $z_1, z_2 \in \mathbb{D}$ is:
\begin{equation}
\text{dist}(z_1, z_2) = 2 \text{arctanh} \left( \frac{|z_1 - z_2|}{|1 - z_1 \bar{z}_2|} \right).
\end{equation}

\indent The isometries of $\mathbb{D}$ are M\"obius transformations in $\text{PSU}(1,1)$. These transformations preserve hyperbolic distances but distort Euclidean lengths.

\subsection{Fuchsian Groups and Quotient Surfaces}

\indent A Fuchsian group is a discrete subgroup of $\text{PSL}(2, \mathbb{R})$, the group of orientation-preserving isometries of $\mathbb{H}^2$. Each Fuchsian group $\Gamma$ has a fundamental domain $\mathcal{F} \subset \mathbb{H}^2$ such that translates of $\mathcal{F}$ tile the hyperbolic plane.

\begin{definition}[Fundamental Domain]
A fundamental domain for a Fuchsian group $\Gamma$ is a connected region $\mathcal{F} \subset \mathbb{H}^2$ such that:
\begin{enumerate}
\item $\bigcup_{\gamma \in \Gamma} \gamma(\mathcal{F}) = \mathbb{H}^2$ (the translates cover the plane),
\item $\gamma_1(\mathcal{F})^\circ \cap \gamma_2(\mathcal{F})^\circ = \emptyset$ for $\gamma_1 \neq \gamma_2$ (interiors are disjoint).
\end{enumerate}
\end{definition}

\indent When $\Gamma$ is torsion-free (no elements of finite order except identity), the quotient $\mathbb{H}^2 / \Gamma$ is a closed orientable surface of genus $g \geq 2$. The genus is determined by the hyperbolic area of the fundamental domain via the Gauss-Bonnet theorem:
\begin{equation}
\text{Area}(\mathcal{F}) = 2\pi(2g - 2) = -2\pi\chi,
\end{equation}
where $\chi = 2 - 2g$ is the Euler characteristic.

\subsection{Schl\"afli Symbols and Tessellations}

\begin{definition}[Schl\"afli Symbol]
A $\{p, q\}$ tessellation is a regular tiling where each face is a regular $p$-gon and each vertex has degree $q$.
\end{definition}

\begin{theorem}[Hyperbolic Condition]
A $\{p,q\}$ tessellation exists in the hyperbolic plane if and only if:
\begin{equation}
(p-2)(q-2) > 4.
\end{equation}
Equivalently: $\frac{1}{p} + \frac{1}{q} < \frac{1}{2}$.
\end{theorem}

\indent For our code families:
\begin{itemize}
\item $\{8,3\}$: $(8-2)(3-2) = 6 > 4$.
\item $\{10,3\}$: $(10-2)(3-2) = 8 > 4$.
\item $\{12,3\}$: $(12-2)(3-2) = 10 > 4$.
\end{itemize}

\subsection{Quotient Surfaces and Combinatorial Parameters}

\indent A closed hyperbolic surface $\Sigma_g$ of genus $g \geq 2$ is realized as a quotient $\mathbb{H}^2 / \Gamma$, where $\Gamma$ is a discrete, torsion-free Fuchsian group. The Euler characteristic relates the cell counts:
\begin{equation}
\chi = V - E + F = 2 - 2g.
\end{equation}

\begin{proposition}[Combinatorial Parameters]
For a $\{p,q\}$ tessellation of a genus-$g$ surface, the constraints $qV = 2E = pF$ yield:
\begin{equation}
F = \frac{4q(g-1)}{pq - 2p - 2q}, \quad E = \frac{pF}{2}, \quad V = \frac{pF}{q}.
\end{equation}
\end{proposition}

\begin{example}[Computing Parameters for $\{8,3\}$, Genus 2]
For $\{8,3\}$ on genus-2:
\begin{align}
F &= \frac{4 \cdot 3 \cdot (2-1)}{8 \cdot 3 - 2 \cdot 8 - 2 \cdot 3} = \frac{12}{24 - 16 - 6} = \frac{12}{2} = 6, \\
E &= \frac{8 \cdot 6}{2} = 24, \quad V = \frac{8 \cdot 6}{3} = 16.
\end{align}
Verification: $V - E + F = 16 - 24 + 6 = -2 = 2 - 2(2)$. \checkmark
\end{example}

\subsection{Asymptotic Scaling}

\indent The distinction between Euclidean and hyperbolic geometry affects quantum code parameters. In Euclidean geometry, area grows polynomially with radius ($A \sim r^2$), while in hyperbolic geometry, area grows exponentially ($A \sim e^r$).

\begin{proposition}[Euclidean Scaling]
For surface codes on Euclidean tessellations (e.g., the square lattice $\{4,4\}$):
\begin{equation}
n = \Theta(d^2), \quad k = \Theta(1), \quad \frac{k}{n} = \Theta(1/n) \to 0.
\end{equation}
The encoding rate vanishes as code size increases.
\end{proposition}

\begin{theorem}[Hyperbolic Scaling]
For hyperbolic surface codes with $n$ physical qubits:
\begin{equation}
\frac{k}{n} = \Theta(1) \quad \text{(constant rate)}, \quad d_{\text{emb}} = \Theta(\log n) \quad \text{(logarithmic distance)}.
\end{equation}
The constant-rate, logarithmic-distance tradeoff is characteristic of hyperbolic codes~\cite{breuckmann2016constructions}.
\end{theorem}

\section{Wythoff Construction}
\label{sec:face_coloring}

\indent Color code tilings on the hyperbolic plane can be constructed through Wythoff's kaleidoscopic method~\cite{higgott2024constructions}. A generator point is placed inside the fundamental triangle and perpendicular lines are projected from that point to each of the bounding mirrors. The resulting tilings have 3-colorable faces. The tiling is generated by a triangle group; quotienting by a normal subgroup yields a finite tiling on a closed surface. Each edge inherits its color as the complement of its two adjacent face colors:
\begin{equation}
\text{color}(e) = 3 - \text{color}(f_1) - \text{color}(f_2),
\end{equation}
where $f_1, f_2$ are the faces incident to edge $e$.

\subsection{Triangle Groups}

\begin{definition}[Triangle Group]
The triangle group $\Delta(p,q,r)$ has presentation:
\begin{equation}
\Delta(p,q,r) = \langle a, b, c \mid a^2 = b^2 = c^2 = (ab)^r = (bc)^q = (ca)^p = e \rangle,
\end{equation}
where $a, b, c$ are reflections across the three sides. The fundamental triangle has interior angles $\pi/p$, $\pi/q$, $\pi/r$ at its vertices.
\end{definition}

\indent The sum of interior angles determines the geometry:
\begin{itemize}
\item $\frac{1}{p} + \frac{1}{q} + \frac{1}{r} > 1$: spherical geometry.
\item $\frac{1}{p} + \frac{1}{q} + \frac{1}{r} = 1$: Euclidean geometry.
\item $\frac{1}{p} + \frac{1}{q} + \frac{1}{r} < 1$: hyperbolic geometry.
\end{itemize}

\indent For a $\{p,q\}$ tessellation, we use $\Delta(p,q,2)$ with $r=2$:
\begin{equation}
\Delta(p,q,2) = \langle a, b, c \mid a^2 = b^2 = c^2 = (ab)^2 = (bc)^q = (ca)^p = e \rangle.
\end{equation}

\indent The group acts on the hyperbolic plane when $\frac{1}{p} + \frac{1}{q} + \frac{1}{2} < 1$. For our families with $q = 3$:
\begin{itemize}
\item $\{8,3\}$: $\frac{1}{8} + \frac{1}{3} + \frac{1}{2} = \frac{23}{24} < 1$.
\item $\{10,3\}$: $\frac{1}{10} + \frac{1}{3} + \frac{1}{2} = \frac{14}{15} < 1$.
\item $\{12,3\}$: $\frac{1}{12} + \frac{1}{3} + \frac{1}{2} = \frac{11}{12} < 1$.
\end{itemize}

\begin{remark}[Rotation Subgroup]
For computation, we use the rotation subgroup $\Delta^+(p,q,2) = \langle \alpha, \beta, \gamma \mid \alpha^2 = \beta^q = \gamma^p = \alpha\beta\gamma = e \rangle$. This index-2 subgroup consists of orientation-preserving isometries. The correspondence with the reflection group is $\alpha = ab$, $\beta = bc$, $\gamma = ca$.
\end{remark}

\subsection{Flag Structure and Coset Enumeration}

\indent Our pipeline extracts the tessellation combinatorics from the coset table using the flag structure of the tiling. A flag in the tessellation is an incident triple (vertex, edge, face). The triangle group acts transitively on flags with trivial stabilizer: each coset of the identity corresponds to exactly one flag.

\begin{definition}[Coset Table]
Given a finite-index normal subgroup $N \trianglelefteq \Delta(p,q,2)$, the coset table records the action of generators on cosets. The number of cosets equals the number of flags in the quotient tessellation.
\end{definition}

\indent The tessellation structure emerges from orbits under subgroups of the rotation group $\Delta^+$:
\begin{itemize}
\item Vertices: orbits under $\langle \beta \rangle$ (each orbit has $q$ elements).
\item Edges: orbits under $\langle \alpha \rangle$ (each orbit has 2 elements).
\item Faces: orbits under $\langle \gamma \rangle$ (each orbit has $p$ elements).
\end{itemize}

\indent The Wythoff construction defines the tessellation geometry for a given triangle group quotient, but it does not by itself enumerate which quotients exist. For that, we need the LINS algorithm described in the next section. The construction pipeline is: LINS enumerates finite-index normal subgroups of $\Delta^+(p,3,2)$, each subgroup defines a quotient tessellation, and the Wythoff construction extracts its vertex, edge, and face structure from the coset table.

\section{Code Construction}
\label{sec:code_construction}

\subsection{The LINS Algorithm}

\indent As described in Section~\ref{sec:face_coloring}, the Wythoff construction produces a tessellation from a given triangle group quotient. The Low-Index Normal Subgroups (LINS) algorithm enumerates finite-index normal subgroups of finitely presented groups. We use the GAP implementation~\cite{gap4} with the LINS package~\cite{lins, firth2005} to enumerate subgroups of the rotation group $\Delta^+(p,3,2)$.

\indent The coset table for $N$ encodes the tessellation combinatorics: vertices are orbits under the generator of order 3, edges are orbits under the generator of order 2, and faces are orbits under the generator of order $p$.

\subsection{Comparison with Prior Construction Methods}

\indent Table~\ref{tab:pipeline_comparison} compares our construction pipeline with prior work.

\begin{table}[htbp]
\centering
\caption{Comparison of hyperbolic code construction methods.}
\label{tab:pipeline_comparison}
\resizebox{\columnwidth}{!}{%
\begin{tabular}{@{}lcccccc@{}}
\toprule
 & Fahimniya et al. & Higgott et al. & Ozawa et al. & Mahmoud et al. & Sutcliffe et al. & This work \\
\midrule
Tessellations & $\{8,3\}$ & $\{8,3\}$ & $\{8,3\}$ & $\{8,3\}$, $\{10,3\}$ & $\{8,3\}$ & $\{p,3\}$\rlap{$^*$} \\
Code type & Floquet & Floquet & Floquet (HCF) & Static CSS & Floquet & Floquet \\
Measurement schedule & XX/YY/ZZ & XX/YY/ZZ & XX/ZZ & --- & XX/YY/ZZ & XX/YY/ZZ \\
Construction & Conder & Wythoff & Fahimniya+Higgott & LINS+Bravais & Foster+Higgott & LINS+Wythoff \\
Distributed & No & No & No & No & Yes & Yes \\
\bottomrule
\multicolumn{7}{@{}l}{\scriptsize $^*$\,Polygon order $p \in \{8,10,12\}$.}
\end{tabular}}
\end{table}

\indent Fahimniya et al.~\cite{fahimniya2023faulttolerant} select $\{8,3\}$ lattices from the Conder database~\cite{conder2002trivalent} of symmetric trivalent graphs and filter by girth~8 and valid face 3-colorings. Higgott and Breuckmann~\cite{higgott2024constructions} use the Wythoff construction and provide Stim circuit files but do not publish generator code. Ozawa et al.~\cite{ozawa2025hyperbolic} apply a six-step XX/ZZ-only schedule to $\{8,3\}$ hyperbolic Floquet codes, citing Fahimniya et al.\ and Higgott and Breuckmann for the underlying code construction. Mahmoud et al.~\cite{mahmoud2025systematic} construct hyperbolic surface codes on $\{p,q\}$ tessellations with specific underlying Bravais lattices ($\{4g,4g\}$ and $\{2(2g{+}1),2g{+}1\}$); they simulate $\{8,3\}$ and $\{10,3\}$ hyperbolic surface codes. Sutcliffe et al.~\cite{sutcliffe2025distributed} take $\{8,3\}$ graphs from the Foster census~\cite{foster1988census, foster2020encyclopedia} and adopt the Floquet code construction of Higgott and Breuckmann~\cite{higgott2024constructions}; they evaluate these codes in distributed settings but do not introduce new code families. Our pipeline derives the tessellation structure directly from the Wythoff construction and uses LINS for subgroup enumeration. The result is a parametric generator for arbitrary $\{p,3\}$ families. Table~\ref{tab:code_params} lists the codes produced by this pipeline.

\subsection{Code Parameters}

\indent Table~\ref{tab:code_params} summarizes the codes constructed. For each code, $n = V$ is the number of physical qubits (vertices), $E$ the number of edges, $F$ the number of faces, $g$ the genus, $k = 2g$ the number of logical qubits, and $d_{\text{emb}}$ is the embedded distance~\cite{higgott2024constructions}. For each color $c \in \{R,G,B\}$, the restricted dual lattice $T^*_c$ excludes faces of color $c$. The embedded distance $d_{\text{emb}}$ equals the minimum weight of a homologically non-trivial cycle or co-cycle in any $T^*_c$.

\begin{table}[htbp]
\centering
\caption{Code parameters for hyperbolic Floquet codes constructed in this work. Codes marked $^\dagger$ have $d_{\text{emb}} = 2$; these cannot correct errors and are excluded from threshold estimation but retained as base codes for fine-graining (Section~\ref{sec:fine_graining}).}
\label{tab:code_params}
\begin{tabular}{llrrrrrrr}
\toprule
Family & Code & $n$ & $E$ & $F$ & $g$ & $k$ & $k/n$ & $d_{\text{emb}}$ \\
\midrule
\multirow{7}{*}{$\{8,3\}$}
& H16$_{8,3}^\dagger$ & 16 & 24 & 6 & 2 & 4 & 0.25 & 2 \\
& H32$_{8,3}^\dagger$ & 32 & 48 & 12 & 3 & 6 & 0.19 & 2 \\
& H64$_{8,3}^\dagger$ & 64 & 96 & 24 & 5 & 10 & 0.16 & 2 \\
& H144$_{8,3}$ & 144 & 216 & 54 & 10 & 20 & 0.14 & 4 \\
& H256$_{8,3}$ & 256 & 384 & 96 & 17 & 34 & 0.13 & 4 \\
& H336$_{8,3}$ & 336 & 504 & 126 & 22 & 44 & 0.13 & 4 \\
& H432$_{8,3}$ & 432 & 648 & 162 & 28 & 56 & 0.13 & 4 \\
\midrule
\multirow{4}{*}{$\{10,3\}$}
& H50$_{10,3}^\dagger$ & 50 & 75 & 15 & 6 & 12 & 0.24 & 2 \\
& H120$_{10,3}^\dagger$ & 120 & 180 & 36 & 13 & 26 & 0.22 & 2 \\
& H250$_{10,3}$ & 250 & 375 & 75 & 26 & 52 & 0.21 & 4 \\
& H720$_{10,3}$ & 720 & 1080 & 216 & 73 & 146 & 0.20 & 4 \\
\midrule
\multirow{5}{*}{$\{12,3\}$}
& H48$_{12,3}^\dagger$ & 48 & 72 & 12 & 7 & 14 & 0.29 & 2 \\
& H72$_{12,3}^\dagger$ & 72 & 108 & 18 & 10 & 20 & 0.28 & 2 \\
& H96$_{12,3}^\dagger$ & 96 & 144 & 24 & 13 & 26 & 0.27 & 2 \\
& H168$_{12,3}^\dagger$ & 168 & 252 & 42 & 22 & 44 & 0.26 & 2 \\
& H312$_{12,3}^\dagger$ & 312 & 468 & 78 & 40 & 80 & 0.26 & 2 \\
\bottomrule
\end{tabular}
\end{table}

\subsection{Fine-Graining and Semi-Hyperbolic Codes}
\label{sec:fine_graining}

\subsubsection{The Fine-Graining Procedure}

\indent Hyperbolic Floquet codes achieve constant encoding rate $k/n = \Theta(1)$ but have distance $d_{\text{emb}} = O(\log n)$. This logarithmic scaling limits error suppression at fixed code size. Breuckmann et al.~\cite{breuckmann2017semihyperbolic} introduced a fine-graining procedure for hyperbolic CSS surface codes. The procedure increases distance while preserving the number of logical qubits. Higgott and Breuckmann~\cite{higgott2024constructions} extended this procedure to hyperbolic Floquet codes. The resulting codes are termed semi-hyperbolic because they interpolate between hyperbolic ($d_{\text{emb}} = O(\log n)$) and Euclidean ($d_{\text{emb}} = O(\sqrt{n})$) distance scaling.

\begin{definition}[Fine-Graining Level]
The fine-graining level $\ell$ is the subdivision parameter. Level $\ell = 1$ is the base code with no subdivision. Level $\ell = 2$ subdivides each edge into 2 segments (each triangle into 4). Level $\ell = 3$ subdivides each edge into 3 segments (each triangle into 9).
\end{definition}

\indent The fine-graining procedure subdivides each triangle in the dual lattice into $\ell^2$ smaller triangles by inserting $\ell-1$ vertices along each edge and $(\ell-1)(\ell-2)/2$ interior vertices per face. For a base code with parameters $[[n, k, d_{\text{emb}}]]$, fine-graining with parameter $\ell$ produces a code with:
\begin{equation}
n' = \ell^2 \cdot n, \quad k' = k.
\end{equation}
The vertex count formula follows from the dual lattice~\cite{higgott2024constructions}. The dual $\mathcal{T}^*$ of a 3-valent color code tiling $\mathcal{T}$ has all-triangular faces, with $|F^*| = |V| = n$ faces (one per vertex of $\mathcal{T}$). Fine-graining replaces each triangular face of $\mathcal{T}^*$ with $\ell^2$ sub-triangles, so $\mathcal{T}_\ell^*$ has $\ell^2 \cdot n$ faces and its dual $\mathcal{T}_\ell$ has $n' = \ell^2 \cdot n$ vertices.

\indent The genus and number of logical qubits remain unchanged because fine-graining is a topological refinement that preserves the surface structure. The embedded distance of each fine-grained code must be computed independently; no universal formula $d_{\text{emb}}'(\ell)$ holds across all codes.

\subsubsection{Geometric Construction}

\indent For codes derived from hyperbolic tessellations, fine-graining must respect the hyperbolic metric to maintain the 3-coloring required for the Floquet measurement schedule. New vertices along edges are placed at hyperbolic geodesic midpoints using M\"obius transformations in the Poincar\'e disk model.

\indent Given two points $z_1, z_2 \in \mathbb{D}$, the geodesic midpoint is computed by:
\begin{enumerate}
\item Map $z_1$ to the origin via the isometry $\phi(z) = (z - z_1)/(1 - \bar{z}_1 z)$. Since $\phi$ is an isometry sending $z_1 \mapsto 0$, the geodesic from $z_1$ to $z_2$ maps to a radial geodesic from $0$ to $\phi(z_2)$.
\item Find the hyperbolic midpoint along this geodesic: $w = \phi(z_2)/(1 + \sqrt{1 - |\phi(z_2)|^2})$.
\item Map back via $\phi^{-1}(w) = (w + z_1)/(1 + \bar{z}_1 w)$.
\end{enumerate}

\indent Interior vertices are placed using hyperbolic barycentric interpolation within each triangle.

\subsubsection{Color Inheritance}

\indent The 3-coloring of faces (equivalently, edges) must extend consistently to the fine-grained lattice. The subdivision of each triangular cell in the dual lattice inherits colors from its three corner vertices. These corners have distinct colors from the base tessellation's face 3-coloring. New vertices introduced by subdivision receive colors such that adjacent vertices always have different colors. This is achieved by assigning colors via linear interpolation in barycentric coordinates modulo 3. The resulting edge coloring partitions edges into three disjoint classes.

\subsubsection{Comparison: Hyperbolic vs Semi-Hyperbolic}

\indent Table~\ref{tab:fine_grained_codes} compares base codes with their fine-grained variants.

\begin{table}[htbp]
\centering
\caption{Base codes and fine-grained variants. Fine-graining increases $n$ and $d_{\text{emb}}$ while preserving $k$.}
\label{tab:fine_grained_codes}
\begin{tabular}{llrrrrl}
\toprule
Base & Level & $n$ & $k$ & $d_{\text{emb}}$ & $k/n$ & Type \\
\midrule
H16$_{8,3}$ & $\ell=1$ & 16 & 4 & 2 & 0.25 & Hyperbolic \\
H16$_{8,3}$ & $\ell=2$ & 64 & 4 & 3 & 0.06 & Semi-hyperbolic \\
H16$_{8,3}$ & $\ell=3$ & 144 & 4 & 4 & 0.03 & Semi-hyperbolic \\
H16$_{8,3}$ & $\ell=4$ & 256 & 4 & 6 & 0.02 & Semi-hyperbolic \\
H16$_{8,3}$ & $\ell=5$ & 400 & 4 & 7 & 0.01 & Semi-hyperbolic \\
\midrule
H64$_{8,3}$ & $\ell=1$ & 64 & 10 & 2 & 0.16 & Hyperbolic \\
H64$_{8,3}$ & $\ell=2$ & 256 & 10 & 4 & 0.04 & Semi-hyperbolic \\
H64$_{8,3}$ & $\ell=3$ & 576 & 10 & 6 & 0.02 & Semi-hyperbolic \\
H64$_{8,3}$ & $\ell=4$ & 1024 & 10 & 10 & 0.01 & Semi-hyperbolic \\
\midrule
H50$_{10,3}$ & $\ell=1$ & 50 & 12 & 2 & 0.24 & Hyperbolic \\
H50$_{10,3}$ & $\ell=2$ & 200 & 12 & 4 & 0.06 & Semi-hyperbolic \\
H50$_{10,3}$ & $\ell=3$ & 450 & 12 & 6 & 0.03 & Semi-hyperbolic \\
H50$_{10,3}$ & $\ell=4$ & 800 & 12 & 8 & 0.02 & Semi-hyperbolic \\
\midrule
H48$_{12,3}$ & $\ell=1$ & 48 & 14 & 2 & 0.29 & Hyperbolic \\
H48$_{12,3}$ & $\ell=2$ & 192 & 14 & 4 & 0.07 & Semi-hyperbolic \\
H48$_{12,3}$ & $\ell=3$ & 432 & 14 & 4 & 0.03 & Semi-hyperbolic \\
H48$_{12,3}$ & $\ell=4$ & 768 & 14 & 7 & 0.02 & Semi-hyperbolic \\
\bottomrule
\end{tabular}
\end{table}

\indent Fine-graining improves distance at the cost of encoding rate. A base H16$_{8,3}$ code with $k/n = 0.25$ becomes a semi-hyperbolic code with $k/n = 0.02$ at $\ell = 4$, but distance increases from 2 to 6. In the distributed setting (Section~\ref{sec:distributed}), fine-graining is essential for achieving positive thresholds.

\section{Distributed Quantum Error Correction}
\label{sec:distributed}

\indent This section evaluates our $\{8,3\}$, $\{10,3\}$, and $\{12,3\}$ codes in the distributed setting introduced in Section~\ref{sec:introduction}. We partition each code across QPUs via spectral bisection and assign local and non-local error rates to edges according to the partition.

\subsection{Distributed Noise Model}

\indent In a distributed architecture, edges of the code graph partition into local edges (both endpoints on the same QPU) and non-local edges (endpoints on different QPUs). We use recursive spectral bisection~\cite{simon1991partitioning} to minimize the edge cut (number of non-local edges). Each QPU holds at most 21 data qubits, as in~\cite{sutcliffe2025distributed}. Spectral bisection achieves non-local edge fractions of 25--37\% across our codes.

\indent We simulate circuits using Stim~\cite{gidney2021stim} and decode using PyMatching~\cite{higgott2023pymatching, higgott2025sparseblossom}. Following the circuit-level noise model of~\cite{sutcliffe2025distributed}, we assign:
\begin{itemize}
\item Local error rate $p_{\text{local}} = 0.0003$ (99.97\% fidelity) for single-qubit gates, two-qubit gates, state preparation, and measurement.
\item Non-local error rate $p_{\text{NL}}$ (swept from 0.1\% to 10\%) modeling Bell state infidelity for inter-QPU measurements.
\item Idle noise: $p_{\text{local}}$ per gate cycle while waiting for Bell state generation (5 cycles assumed).
\end{itemize}

\subsection{Error Rate Metrics}
\label{sec:metrics}

\indent We use two logical error rate metrics (Table~\ref{tab:metric_map}) and two threshold definitions (Table~\ref{tab:threshold_defs}). The \emph{any-logical} metric is the probability that any of the $k$ encoded logical qubits is corrupted by an uncorrected error. The \emph{worst-case single logical} metric $\epsilon_L$ is the maximum per-round error rate across the $k$ logical $Z$ observables.

\begin{table}[htbp]
\centering
\caption{Logical error rate metrics.}
\label{tab:metric_map}
\begin{tabular}{@{}lll@{}}
\toprule
Metric & Definition & Used in \\
\midrule
Any-logical & Failure if any of $k$ logicals is corrupted &
Table~\ref{tab:sdem3_any_logical} \\
Worst-case $\epsilon_L$ & Max per-round $\epsilon_L$ across $k$ logical $Z$ observables &
Table~\ref{tab:distributed_sutcliffe} \\
\bottomrule
\end{tabular}
\end{table}

\begin{table}[htbp]
\centering
\caption{Threshold definitions.}
\label{tab:threshold_defs}
\begin{tabular}{@{}lll@{}}
\toprule
Threshold & Definition & Used in \\
\midrule
Pseudo-threshold $\tilde{p}_{\text{NL}}$ & Highest swept $p_{\text{NL}}$ at which $\epsilon_L \leq p_{\text{local}}$ for a single code &
Tables~\ref{tab:sdem3_distributed}--\ref{tab:comprehensive_summary} \\
Crossing threshold $p_{\text{NL}}^*$ & $p_{\text{NL}}$ where same-$k$ codes cross in per-observable rate &
Table~\ref{tab:literature_comparison} \\
\bottomrule
\end{tabular}
\end{table}

\indent Per-round rates $\epsilon_L$ and total error rates $p_{\text{total}}$ are related by $p_{\text{total}} = \frac{1}{2}[1 - (1 - 2\epsilon_L)^{n_{\text{rounds}}}]$ (Equation~2 of~\cite{sutcliffe2025distributed}). We compute $\epsilon_L$ for each of the $k$ logical $Z$ observables individually and take the maximum.

\indent A \emph{sub-round} (or color round) measures all edges of a single color. A \emph{plaquette-forming round} consists of three sub-rounds (one each of $XX$, $YY$, $ZZ$). A \emph{detector round} consists of two plaquette-forming rounds~\cite{sutcliffe2025distributed}. We simulate 12 detector rounds throughout.

\subsection{Results on Our Code Families}

\indent Table~\ref{tab:distributed_sutcliffe} presents worst-case single logical $Z$ error rates (Section~\ref{sec:metrics}) at $p_{\text{NL}} = 1\%$ for direct comparison with~\cite{sutcliffe2025distributed}. At $p_{\text{NL}} = 1\%$, our H64$_{8,3}$ achieves $\epsilon_L = 10.7\%$ per detector round. All subsequent sections report non-local pseudo-thresholds derived from $\epsilon_L$ (Section~\ref{sec:metrics}).

\begin{table}[htbp]
\centering
\caption{Worst-case single logical $Z$ error rate $\epsilon_L$ (\% per detector round) under the distributed depolarizing model of~\cite{sutcliffe2025distributed}. NL denotes non-local edge fraction. Simulated with $p_{\text{local}} = 0.03\%$, $p_{\text{NL}} = 1\%$, 12 detector rounds, 5000 shots, 21 data qubits per QPU.}
\label{tab:distributed_sutcliffe}
\begin{tabular}{lrrrr|r}
\toprule
Family & $n$ & $k$ & NL & QPUs & $\epsilon_L$ (\%/round) \\
\midrule
\multirow{4}{*}{$\{8,3\}$}
& 64 & 10 & 33\% & 4 & 10.7 \\
& 144 & 20 & 37\% & 10 & 12.0 \\
& 256 & 34 & 35\% & 17 & 14.3 \\
& 336 & 44 & 36\% & 24 & 22.4 \\
\midrule
\multirow{2}{*}{$\{10,3\}$}
& 50 & 12 & 29\% & 4 & 7.8 \\
& 120 & 26 & 35\% & 8 & 13.5 \\
\midrule
{$\{12,3\}$}
& 72 & 20 & 26\% & 4 & 10.6 \\
\bottomrule
\end{tabular}
\end{table}

\subsection{Distributed SDEM3}
\label{sec:sdem3_distributed}

\indent We write $p_{\text{phys}}$ for the per-edge physical error rate, equal to $p_{\text{NL}}$ on non-local edges and $p_{\text{local}}$ on local edges. For the SDEM3 (Single-step Depolarizing EM3) noise model~\cite{higgott2024constructions}, we apply:
\begin{itemize}
\item Two-qubit depolarizing noise with strength $15p_{\text{phys}}/16$ before each MPP.
\item Measurement outcome flip with probability $p_{\text{phys}}/2$ per edge.
\item Bit-flip noise with probability $p_{\text{phys}}/2$ after initialization and before final measurement.
\item No idle noise.
\end{itemize}

\indent To isolate the effect of measurement outcome noise, we compare SDEM3 against the distributed depolarizing model from the preceding sections. We also include an intermediate model (ancilla-EM3) that adds measurement flips to distributed depolarizing.

\begin{enumerate}
\item \emph{Distributed depolarizing}: Two-qubit depolarizing noise with strength $p_{\text{phys}}$ per edge, idle depolarizing noise with strength $p_{\text{idle}}$ for $t=5$ gate cycles during Bell wait, no measurement outcome noise. This is the model used in the preceding sections.
\item \emph{ancilla-EM3} (intermediate): As above, plus measurement outcome flip with probability $p_{\text{phys}}/2$ per edge.
\item \emph{SDEM3}: The native pair-measurement model applied to the distributed setting, with $p_{\text{phys}} = p_{\text{NL}}$ for non-local edges and $p_{\text{phys}} = p_{\text{local}}$ for local edges. Bell wait idle noise is retained, but H-gate noise and per-round idle noise on measured qubits are removed.
\end{enumerate}

\indent Table~\ref{tab:sdem3_distributed} compares the three models across all three tessellation families at $p_{\text{local}} = 0.03\%$.

\begin{table}[htbp]
\centering
\caption{Distributed pseudo-threshold $\tilde{p}_{\text{NL}}$ (\%) under three noise models at $p_{\text{local}} = 0.03\%$, 12 detector rounds, 3000 shots. NL denotes non-local edge fraction. Entries marked ``---'' indicate $\tilde{p}_{\text{NL}} < 0.5\%$.}
\label{tab:sdem3_distributed}
\begin{tabular}{lrrrrrrr}
\toprule
Code & $n$ & $k$ & NL & Dist.\ depol. & Anc-EM3 & SDEM3 \\
\midrule
H16$_{8,3}$-f3 & 144 & 4 & 27\% & 1.25 & 0.50 & 0.50 \\
H16$_{8,3}$-f4 & 256 & 4 & 25\% & 2.00 & 1.00 & 1.00 \\
H16$_{8,3}$-f5 & 400 & 4 & 28\% & 2.50 & 1.25 & 1.25 \\
H64$_{8,3}$-f3 & 576 & 10 & 28\% & 2.00 & 1.25 & 1.25 \\
H64$_{8,3}$-f4 & 1024 & 10 & 29\% & 3.00 & 1.75 & 1.75 \\
H144$_{8,3}$-f3 & 1296 & 20 & 29\% & 3.00 & 1.50 & 1.50 \\
\midrule
H50$_{10,3}$-f3 & 450 & 12 & 28\% & 2.00 & 0.75 & 0.75 \\
H50$_{10,3}$-f4 & 800 & 12 & 28\% & 3.00 & 1.25 & 1.25 \\
H120$_{10,3}$-f3 & 1080 & 26 & 29\% & 2.00 & 1.00 & 1.00 \\
\midrule
H48$_{12,3}$-f3 & 432 & 14 & 27\% & 1.25 & 0.50 & --- \\
H48$_{12,3}$-f4 & 768 & 14 & 28\% & 1.75 & 1.00 & 1.00 \\
H72$_{12,3}$-f3 & 648 & 20 & 29\% & 1.50 & 0.75 & 0.75 \\
H96$_{12,3}$-f3 & 864 & 26 & 29\% & 1.75 & 0.75 & 0.75 \\
\bottomrule
\end{tabular}
\end{table}

\indent Adding measurement outcome noise reduces pseudo-thresholds by roughly a factor of two across all codes. The $\{8,3\}$ family retains ancilla-EM3 pseudo-thresholds of 0.50--1.75\%, while $\{10,3\}$ achieves up to 1.25\% and $\{12,3\}$ up to 1.00\%. Table~\ref{tab:sdem3_any_logical} shows any-logical error rates across the three models at selected $p_{\text{NL}}$ values for H144$_{8,3}$-f3.

\begin{table}[htbp]
\centering
\caption{Any-logical error rate (\%) for H144$_{8,3}$-f3 ($[[1296,20]]$, 90 QPUs) under three noise models in the distributed setting. 3000 shots, 12 detector rounds, $p_{\text{local}} = 0.03\%$.}
\label{tab:sdem3_any_logical}
\begin{tabular}{rrrr}
\toprule
$p_{\text{NL}}$ & Dist.\ depol. & Anc-EM3 & SDEM3 \\
\midrule
1.5\% & 0.0 & 0.8 & 0.4 \\
2.0\% & 0.0 & 4.5 & 3.9 \\
2.5\% & 0.1 & 24.1 & 19.3 \\
3.0\% & 0.6 & 60.1 & 50.9 \\
3.5\% & 1.0 & 91.1 & 85.9 \\
4.0\% & 2.8 & 99.3 & 97.7 \\
\bottomrule
\end{tabular}
\end{table}

\subsection{Distributed Correlated EM3}
\label{sec:correlated_em3_distributed}

\indent In monolithic settings, correlated EM3 achieves ${\sim}1.5$--$2.0\%$ threshold~\cite{higgott2024constructions}. We now evaluate correlated EM3 on our distributed codes.

\subsubsection{Correlated EM3 Distributed Model}

\indent We implement the correlated EM3 model~\cite{higgott2024constructions} with per-edge error rates baked directly into the circuit generator. Each edge measurement uses an ancilla qubit and 31 correlated \texttt{E}$(q)$ instructions with $q = p_{\text{phys}}/32$, where $p_{\text{phys}} = p_{\text{NL}}$ for non-local edges and $p_{\text{phys}} = p_{\text{local}}$ for local edges. The 31 instructions cover all outcomes in $\{I,X,Y,Z\}^{\otimes 2} \times \{\text{flip}, \text{no-flip}\}$ except the trivial $(II, \text{no-flip})$.

\subsubsection{Results}

\indent Table~\ref{tab:em3_correlated_distributed} reports correlated EM3 distributed pseudo-thresholds at both fine-graining levels alongside the distributed depolarizing model for comparison.

\begin{table}[htbp]
\centering
\caption{Distributed pseudo-threshold $\tilde{p}_{\text{NL}}$ (\%) under correlated EM3 noise at $p_{\text{local}} = 0.03\%$, 12 detector rounds, 3000 shots. Entries marked ``---'' indicate $\tilde{p}_{\text{NL}} < 0.5\%$; ``N/C'' indicates not computed (circuit too large for available memory).}
\label{tab:em3_correlated_distributed}
\begin{tabular}{llrrrrrrr}
\toprule
Family & Base & $k$ & \multicolumn{2}{c}{$\ell=2$} & \multicolumn{2}{c}{$\ell=3$} & \multicolumn{2}{c}{$\ell=4$} \\
\cmidrule(lr){4-5} \cmidrule(lr){6-7} \cmidrule(lr){8-9}
 & & & $n$ & $\tilde{p}_{\text{NL}}$ & $n$ & $\tilde{p}_{\text{NL}}$ & $n$ & $\tilde{p}_{\text{NL}}$ \\
\midrule
\multirow{3}{*}{$\{8,3\}$}
& H16$_{8,3}$ & 4 & 64 & --- & 144 & --- & 256 & 0.50 \\
& H64$_{8,3}$ & 10 & 256 & --- & 576 & 0.50 & 1024 & 0.75 \\
& H144$_{8,3}$ & 20 & 576 & 0.50 & 1296 & 0.75 & & N/C \\
\midrule
\multirow{3}{*}{$\{10,3\}$}
& H50$_{10,3}$ & 12 & 200 & --- & 450 & 0.50 & 800 & 0.75 \\
& H120$_{10,3}$ & 26 & 480 & --- & 1080 & 0.50 & & N/C \\
& H250$_{10,3}$ & 52 & 1000 & --- & 2250 & N/C & & N/C \\
\midrule
\multirow{3}{*}{$\{12,3\}$}
& H48$_{12,3}$ & 14 & 192 & --- & 432 & --- & 768 & 0.50 \\
& H72$_{12,3}$ & 20 & 288 & --- & 648 & --- & & N/C \\
& H96$_{12,3}$ & 26 & 384 & --- & 864 & 0.50 & & N/C \\
\bottomrule
\end{tabular}
\end{table}

\indent H250$_{10,3}$-f3 ($n = 2250$, $k = 52$) could not be simulated under correlated EM3 because the 31 \texttt{E} instructions per edge produce circuits exceeding available memory at this scale.

\indent Table~\ref{tab:em3_correlated_comparison} compares all four noise models in the distributed setting.

\begin{table}[htbp]
\centering
\caption{Distributed pseudo-threshold $\tilde{p}_{\text{NL}}$ (\%) under four noise models at $p_{\text{local}} = 0.03\%$, 12 detector rounds, 3000 shots. Both $\ell = 3$ and $\ell = 4$ codes are shown where available. A dash (---) indicates $\tilde{p}_{\text{NL}} < 0.5\%$.}
\label{tab:em3_correlated_comparison}
\begin{tabular}{lrrrcccc}
\toprule
Code & $n$ & $k$ & NL & Dist.\ depol. & Anc-EM3 & SDEM3 & Corr.\ EM3 \\
\midrule
H16$_{8,3}$-f3 & 144 & 4 & 27\% & 1.25 & 0.50 & 0.50 & --- \\
H16$_{8,3}$-f4 & 256 & 4 & 25\% & 2.00 & 1.00 & 1.00 & 0.50 \\
H16$_{8,3}$-f5 & 400 & 4 & 28\% & 2.50 & 1.25 & 1.25 & 0.75 \\
H64$_{8,3}$-f3 & 576 & 10 & 28\% & 2.00 & 1.25 & 1.25 & 0.50 \\
H64$_{8,3}$-f4 & 1024 & 10 & 29\% & 3.00 & 1.75 & 1.75 & 0.75 \\
H144$_{8,3}$-f3 & 1296 & 20 & 29\% & 3.00 & 1.50 & 1.50 & 0.75 \\
\midrule
H50$_{10,3}$-f3 & 450 & 12 & 28\% & 2.00 & 0.75 & 0.75 & 0.50 \\
H50$_{10,3}$-f4 & 800 & 12 & 28\% & 3.00 & 1.25 & 1.25 & 0.75 \\
H120$_{10,3}$-f3 & 1080 & 26 & 29\% & 2.00 & 1.00 & 1.00 & 0.50 \\
\midrule
H48$_{12,3}$-f3 & 432 & 14 & 27\% & 1.25 & 0.50 & --- & --- \\
H48$_{12,3}$-f4 & 768 & 14 & 28\% & 1.75 & 1.00 & 1.00 & 0.50 \\
H72$_{12,3}$-f3 & 648 & 20 & 29\% & 1.50 & 0.75 & 0.75 & --- \\
H96$_{12,3}$-f3 & 864 & 26 & 29\% & 1.75 & 0.75 & 0.75 & 0.50 \\
\bottomrule
\end{tabular}
\end{table}

\subsection{Distributed Erasure}
\label{sec:distributed_erasure}

\indent The preceding sections parameterize noise by depolarizing error rates ($p_{\text{NL}}$, $p_{\text{local}}$). In spin-optical hardware, photon loss is the primary source of errors~\cite{dessertaine2024enhanced}, so this section switches to erasure rates ($\varepsilon_{\text{NL}}$, $\varepsilon_{\text{local}}$).

\subsubsection{Erasure Noise Model}

\indent We adopt the photon-loss model of Dessertaine et al.~\cite{dessertaine2024enhanced}. Each pair measurement independently fails with probability $p_{\text{RUS}}$ due to a failed repeat-until-success (RUS) protocol. A failed measurement is detected: the decoder knows which measurements were erased. The single-photon loss rate $\varepsilon$ and the RUS failure probability are related by
\begin{equation}
p_{\text{RUS}} = \frac{2 - 2(1-\varepsilon)^2}{2 - (1-\varepsilon)^2}.
\end{equation}

\subsubsection{Distributed Erasure Model}

\indent Dessertaine et al.~\cite{dessertaine2024enhanced} analyze a monolithic spin-optical architecture with uniform photon loss rate $\varepsilon$. We extend this to the distributed setting: local measurements (both qubits on the same QPU) have loss rate $\varepsilon_{\text{local}}$, while non-local measurements (qubits on different QPUs) have loss rate $\varepsilon_{\text{NL}} > \varepsilon_{\text{local}}$. Partitioning follows Section~\ref{sec:distributed}.

\indent Table~\ref{tab:erasure_distributed} reports distributed erasure results for the $\{8,3\}$ Bolza family with $\varepsilon_{\text{local}} = 1\%$. Table~\ref{tab:erasure_dist_cross} extends the comparison across all four same-$k$ sub-families with three or more codes. Each sub-family contains codes with the same $k$ so that the any-logical metric is consistent across code sizes. Based on 5\% sweep resolution, distributed erasure thresholds lie between 35\% and 40\% for $\{8,3\}$ ($k = 4$ and $k = 10$), between 30\% and 35\% for $\{10,3\}$ ($k = 12$), and between 25\% and 30\% for $\{12,3\}$ ($k = 14$).

\begin{table}[htbp]
\centering
\caption{Distributed erasure any-logical error rate (\%) for $\{8,3\}$ Bolza codes ($k = 4$), $\varepsilon_{\text{local}} = 1\%$, varying $\varepsilon_{\text{NL}}$. $M = 1000$ instances, $N = 128$ shots, 21 qubits per QPU.}
\label{tab:erasure_distributed}
\begin{tabular}{lrrrccccc}
\toprule
Code & $n$ & QPUs & NL\% & 15\% & 20\% & 25\% & 30\% & 35\% \\
\midrule
H16$_{8,3}$-f3 & 144 & 8 & 27\% & 2.5 & 10.7 & 23.5 & 39.4 & 47.7 \\
H16$_{8,3}$-f4 & 256 & 16 & 25\% & 0.5 & 3.2 & 16.3 & 37.5 & 47.7 \\
H16$_{8,3}$-f5 & 400 & 28 & 28\% & 0.1 & 0.6 & 4.9 & 25.6 & 45.2 \\
\bottomrule
\end{tabular}
\end{table}

\begin{table}[htbp]
\centering
\caption{Distributed erasure any-logical error rate (\%) across tessellation families, $\varepsilon_{\text{local}} = 1\%$, $M = 1000$ instances, $N = 128$ shots. Each sub-family compares codes with the same $k$, so that the any-logical metric is consistent across code sizes within each group.}
\label{tab:erasure_dist_cross}
\begin{tabular}{lrrrccccc}
\toprule
Code & $n$ & QPUs & NL\% & 10\% & 15\% & 20\% & 25\% & 30\% \\
\midrule
\multicolumn{9}{l}{\emph{$\{8,3\}$ ($k = 4$)}} \\
H16$_{8,3}$-f3 & 144 & 8 & 27\% & 0.4 & 2.5 & 10.7 & 23.5 & 39.4 \\
H16$_{8,3}$-f4 & 256 & 16 & 25\% & 0.0 & 0.5 & 3.2 & 16.3 & 37.5 \\
H16$_{8,3}$-f5 & 400 & 28 & 28\% & 0.0 & 0.1 & 0.6 & 4.9 & 25.6 \\
\midrule
\multicolumn{9}{l}{\emph{$\{8,3\}$ ($k = 10$)}} \\
H64$_{8,3}$-f2 & 256 & 17 & 31\% & 0.1 & 1.4 & 8.7 & 30.8 & 47.3 \\
H64$_{8,3}$-f3 & 576 & 38 & 28\% & 0.0 & 0.1 & 0.6 & 11.2 & 39.2 \\
H64$_{8,3}$-f4 & 1024 & 71 & 29\% & 0.0 & 0.0 & 0.1 & 2.7 & 28.7 \\
\midrule
\multicolumn{9}{l}{\emph{$\{10,3\}$ ($k = 12$)}} \\
H50$_{10,3}$-f2 & 200 & 16 & 32\% & 0.5 & 8.6 & 31.5 & 48.0 & 50.0 \\
H50$_{10,3}$-f3 & 450 & 33 & 28\% & 0.1 & 0.6 & 8.7 & 34.5 & 49.6 \\
H50$_{10,3}$-f4 & 800 & 56 & 28\% & 0.0 & 0.0 & 0.8 & 11.0 & 45.3 \\
\midrule
\multicolumn{9}{l}{\emph{$\{12,3\}$ ($k = 14$)}} \\
H48$_{12,3}$-f2 & 192 & 14 & 30\% & 14.4 & 38.5 & 49.3 & 50.0 & 50.0 \\
H48$_{12,3}$-f3 & 432 & 30 & 27\% & 1.3 & 12.5 & 34.8 & 49.2 & 50.0 \\
H48$_{12,3}$-f4 & 768 & 54 & 28\% & 0.1 & 2.6 & 12.0 & 35.2 & 49.9 \\
\bottomrule
\end{tabular}
\end{table}

\subsection{Summary and Comparison}
\label{sec:summary}

\indent Table~\ref{tab:comprehensive_summary} consolidates distributed pseudo-threshold results across all noise models, code families, and fine-graining levels. All depolarizing pseudo-thresholds use $p_{\text{local}} = 0.03\%$; erasure uses $\varepsilon_{\text{local}} = 1\%$.

\begin{table}[htbp]
\centering
\caption{Distributed pseudo-threshold $\tilde{p}_{\text{NL}}$ (\%) across noise models. Both $\ell = 3$ and $\ell = 4$ codes are shown where available. A dash (---) indicates pseudo-threshold below sweep resolution ($< 0.5\%$). Depolarizing pseudo-thresholds have $\pm 0.25\%$ precision (0.5\% sweep intervals); erasure thresholds (in $\varepsilon_{\text{NL}}$ \%) have $\pm 2.5\%$ precision (5\% sweep intervals) and are determined from same-$k$ sub-families (Table~\ref{tab:erasure_dist_cross}). Anc-EM3 is an intermediate comparison (see footnote in Section~\ref{sec:sdem3_distributed}). All depolarizing models use $p_{\text{local}} = 0.03\%$; erasure uses $\varepsilon_{\text{local}} = 1\%$.}
\label{tab:comprehensive_summary}
\begin{tabular}{lrrccccc}
\toprule
 & & & \multicolumn{4}{c}{$\tilde{p}_{\text{NL}}$ (\%)} & $\varepsilon_{\text{NL}}^*$ (\%) \\
\cmidrule(lr){4-7} \cmidrule(lr){8-8}
Code & $n$ & $k$ & Dist.\ depol. & Anc-EM3 & SDEM3 & Corr.\ EM3 & Erasure \\
\midrule
H16$_{8,3}$-f3 & 144 & 4 & 1.25 & 0.50 & 0.50 & --- & 35--40 \\
H16$_{8,3}$-f4 & 256 & 4 & 2.00 & 1.00 & 1.00 & 0.50 & 35--40 \\
H16$_{8,3}$-f5 & 400 & 4 & 2.50 & 1.25 & 1.25 & 0.75 & 35--40 \\
H64$_{8,3}$-f3 & 576 & 10 & 2.00 & 1.25 & 1.25 & 0.50 & 35--40 \\
H64$_{8,3}$-f4 & 1024 & 10 & 3.00 & 1.75 & 1.75 & 0.75 & 35--40 \\
H144$_{8,3}$-f3 & 1296 & 20 & 3.00 & 1.50 & 1.50 & 0.75 & 35--40 \\
\midrule
H50$_{10,3}$-f3 & 450 & 12 & 2.00 & 0.75 & 0.75 & 0.50 & 30--35 \\
H50$_{10,3}$-f4 & 800 & 12 & 3.00 & 1.25 & 1.25 & 0.75 & 30--35 \\
H120$_{10,3}$-f3 & 1080 & 26 & 2.00 & 1.00 & 1.00 & 0.50 & 30--35 \\
\midrule
H48$_{12,3}$-f3 & 432 & 14 & 1.25 & 0.50 & --- & --- & 25--30 \\
H48$_{12,3}$-f4 & 768 & 14 & 1.75 & 1.00 & 1.00 & 0.50 & 25--30 \\
H72$_{12,3}$-f3 & 648 & 20 & 1.50 & 0.75 & 0.75 & --- & 25--30 \\
H96$_{12,3}$-f3 & 864 & 26 & 1.75 & 0.75 & 0.75 & 0.50 & 25--30 \\
\bottomrule
\end{tabular}
\end{table}

\begin{table}[htbp]
\centering
\caption{Comparison with literature. $\tilde{p}_{\text{NL}}$: pseudo-threshold (maximum across codes in family). $p_{\text{NL}}^*$: crossing-based threshold from same-$k$ families with $\ell = 2, 3, 4$ fine-graining levels.}
\label{tab:literature_comparison}
\begin{tabular}{llccccc}
\toprule
Code family & Reference & Noise model & $\ell$ & $\tilde{p}_{\text{NL}}$ & $p_{\text{NL}}^*$ & $k$ \\
\midrule
Hyp.\ Floquet $\{8,3\}$ & Sutcliffe et al.~\cite{sutcliffe2025distributed} & Dist.\ depol. & 3 & $3.1\%$ & --- & 4 \\
\midrule
Hyp.\ $\{8,3\}$ (this work) & & Dist.\ depol. & 2--4 & $3.0\%$ & ${>}5\%$ & 4--20 \\
Hyp.\ $\{8,3\}$ (this work) & & Corr.\ EM3 & 2--4 & $0.75\%$ & ${\sim}2.3$--$2.9\%$ & 4--20 \\
Hyp.\ $\{8,3\}$ (this work) & & SDEM3 & 2--4 & $1.75\%$ & ${\sim}4$--$5\%$ & 4--20 \\
\midrule
Hyp.\ $\{10,3\}$ (this work) & & Dist.\ depol. & 2--4 & $3.0\%$ & ${>}5\%$ & 12--26 \\
Hyp.\ $\{10,3\}$ (this work) & & Corr.\ EM3 & 2--4 & $0.75\%$ & ${\sim}2.0$--$2.5\%$ & 12--26 \\
Hyp.\ $\{10,3\}$ (this work) & & SDEM3 & 2--4 & $1.25\%$ & ${\sim}4$--$5\%$ & 12--26 \\
\midrule
Hyp.\ $\{12,3\}$ (this work) & & Dist.\ depol. & 2--4 & $1.75\%$ & ${>}5\%$ & 14--26 \\
Hyp.\ $\{12,3\}$ (this work) & & Corr.\ EM3 & 2--4 & $0.50\%$ & ${\sim}1.75$--$2.0\%$ & 14--26 \\
Hyp.\ $\{12,3\}$ (this work) & & SDEM3 & 2--4 & $1.00\%$ & ${\sim}4$--$5\%$ & 14--26 \\
\bottomrule
\end{tabular}
\end{table}

\indent Table~\ref{tab:literature_comparison} compares with prior work. Sutcliffe et al.~\cite{sutcliffe2025distributed} report a pseudo-threshold of 3.1\% for their largest semi-hyperbolic $\{8,3\}$ code. Tables~\ref{tab:sdem3_distributed}--\ref{tab:comprehensive_summary} report pseudo-thresholds computed by the same method (Section~\ref{sec:metrics}). Our largest $\{8,3\}$ depolarizing pseudo-threshold of 3.0\% uses comparable parameters to Sutcliffe et al.: $p_{\text{local}} = 0.03\%$, 21 data qubits per QPU, spectral bisection partitioning, and the same noise model (Table~\ref{tab:distributed_sutcliffe}).

\indent To verify that error suppression improves with code size, we also compute crossing-based thresholds by comparing per-observable error rates across codes with the same $k$ within each family ($\ell = 2, 3, 4$ fine-graining levels). Under correlated EM3, crossing thresholds are ${\sim}2.3$--$2.9\%$ for $k = 4$ $\{8,3\}$, ${\sim}2.5\%$ for $k = 10$ $\{8,3\}$, ${\sim}2.5\%$ for $k = 12$ $\{10,3\}$, and ${\sim}1.75$--$2.0\%$ for $k = 14$ $\{12,3\}$. Under distributed depolarizing and ancilla-EM3, crossing thresholds exceed the $5\%$ upper limit of our sweep range for all families; larger codes outperform smaller ones throughout the measured range. Under SDEM3, crossing thresholds are ${\sim}4$--$5\%$ for families with three or more codes.

\clearpage
\section{Conclusion}
\label{sec:conclusion}

\indent We constructed hyperbolic Floquet codes from $\{8,3\}$, $\{10,3\}$, and $\{12,3\}$ tessellations using the LINS algorithm with the Wythoff kaleidoscopic construction and distributed them across QPUs via spectral bisection. The $\{10,3\}$ and $\{12,3\}$ families are new to hyperbolic Floquet codes; prior work focused on $\{8,3\}$ tessellations. The $\{8,3\}$ family achieves the highest pseudo-thresholds across all noise models. Its asymptotic genus-per-vertex ratio of 6.25\% is the lowest of the three families ($\{10,3\}$: 10.0\%, $\{12,3\}$: 12.5\%). Fewer logical qubits per physical qubit means more physical resources per logical qubit and better error suppression.

\indent This paper studied four noise models: depolarizing, SDEM3, correlated EM3, and erasure. For depolarizing noise, fine-grained codes achieve distributed pseudo-thresholds up to 3.0\% for $\{8,3\}$ and $\{10,3\}$, and 1.75\% for $\{12,3\}$. The $\{8,3\}$ depolarizing pseudo-threshold of 3.0\% corresponds to a non-local fidelity of 97.0\% at local fidelity 99.97\%. Under SDEM3, pseudo-thresholds reach 1.75\% for $\{8,3\}$, 1.25\% for $\{10,3\}$, and 1.00\% for $\{12,3\}$. Under correlated EM3~\cite{higgott2024constructions}, same-$k$ code families yield crossing-based thresholds of ${\sim}1.75$--$2.9\%$ across all three tessellations. Under spin-optical erasure noise~\cite{dessertaine2024enhanced} at 1\% local loss, thresholds reach 35--40\% for $\{8,3\}$, 30--35\% for $\{10,3\}$, and 25--30\% for $\{12,3\}$.

\indent Across all noise models, the non-local pseudo-thresholds of standard hyperbolic Floquet codes exceed local error rates by one to two orders of magnitude. Distributed hyperbolic Floquet codes offer a practical route to scalable quantum error correction across networked QPUs.

\subsection{Future Directions}

\begin{enumerate}
\item Seam thresholds: distributed surface codes exhibit elevated thresholds for seam stabilizers spanning QPU boundaries~\cite{ramette2024faulttolerant}. Whether distributed hyperbolic codes show a similar seam-bulk threshold separation is unknown.

\item Multi-parameter noise analysis: our erasure and depolarizing thresholds are single-parameter sweeps. Dessertaine et al.~\cite{dessertaine2024enhanced} compute a fault-tolerant region by sweeping photon loss, distinguishability, and decoherence simultaneously. Extending this multi-parameter analysis to distributed hyperbolic codes would enable direct comparison of fault-tolerant region volumes.
\item Logical gates: Lattice surgery~\cite{horsman2012lattice} has been developed for planar surface codes, and Dehn twists have been proposed for hyperbolic surface codes~\cite{lavasani2019universal}; adapting these techniques to the distributed hyperbolic Floquet setting would enable fault-tolerant computation.
\end{enumerate}


\end{document}